# A Novel Approach for Lattice Simulations of Polymer Chains in Dense Amorphous Polymer Systems: Method Development and Validation with 2-D Lattices

Jaydeep A. Kulkarni[1]
Joydeep Mukherjee[2]
Ryan C. Snyder[3]
Timothy W. King[4]
Antony N. Beris[5,*]

[1]Fluent Inc, 500 Davis Street, Suite 600, Evanston IL 60201
[2] The Dow Chemical Company, Freeport, TX 77541
[3]Department of Chemical Engineering, University of California, Santa Barbara, CA 93106
[4]Department of Chemical Engineering, University of Virginia, Charlottesville, VA 21904
[5]Department of Chemical Engineering, University of Delaware, Newark, DE 19716; Email : beris@udel.edu, Phone: +1(302)831 8018, Fax: +1(302)831 1048
[*]Author to whom correspondence should be addressed

## Summary

We present here the systematic development of quantitative lattice simulations of dense polymers through a novel computational technique that allows for an efficient accounting of the chain conformations. Our approach is based on the decomposition of the original lattice into sublattices of optimal size. We develop and validate the method here for 2-D lattices using sublattices of 4x4 nodes. For each possible connectivity, i.e. arrangement of bonds connecting the 4x4 nodes of a sublattice with the rest of the nodes of the lattice, all possible sublattice microstates (submicrostates) are evaluated. This information, as well as other related information pertaining to the sublattices, needed later on in the algorithm, is generated separately and stored in databases. All that information is subsequently used for the efficient generation of the microstates of the full lattice. We apply this technique to study the interlamellar amorphous phase in dense semicrystalline polymers where in polymer chains conform to a 2-D square lattice.

For lattices of moderate size (up to 8x8 nodes), exact results can be obtained from an exhaustive enumeration of all the microstates corresponding to the fully dense (i.e. with no free chain ends) interlamellar amorphous phase of a semicrystalline system. For larger lattices, a stochastic enumeration technique (purely entropic) and an efficient Metropolis Monte Carlo scheme were developed. A large selection of Monte Carlo moves makes the correlation between the Monte Carlo moves especially short. Thus, statistical quantities of interest can be obtained with tight error bars (calculated concurrently with the averages) using small number of steps. The ergodicity of the scheme was verified for small lattices and the results from the Monte Carlo scheme were validated against those from the exhaustive enumeration for lattices of moderate sizes.

## 1 Introduction

Lattice models have been widely used to investigate mesoscopic morphology of semicrystalline polymers[1-20]. Yoon and Flory[2] were the first to use them to study the adjacent and irregular re-entry of chains from the crystalline to the amorphous region and, in particular, the scattering behavior in lamellar semicrystalline polymers. Since then, lattice models have been used extensively - see, for example Kumar and Yoon[3] and Kulkarni and Beris[4] for a broader review. Various approaches have been used for the study of polymer chains using lattice models that can be summarily classified into three categories: Random walk models based on analytical, purely statistical, accounting of chain configurations[5-9], mean field models based on the formulation and minimization of the free energy of polymer chains[3],[10-15] and computationally intensive models where polymer chains are configured on a lattice accounting for chain-chain excluded volume interactions[16-20] with the potential for handling more complex chain-chain energetic interactions. We shall focus only on the computationally intensive approaches because unlike the other methods, they are devoid of approximation errors pertaining to self-avoiding random walks.

For dense polymer systems the traditional approach has been based on individual chain construction in lattice abiding the constraints of excluded volume interactions. The generation of such polymer chain conformations is computationally very expensive due to the large number of rejections. Even the best Monte Carlo approaches[16-20] are plagued by uncertainties, or, at best, low computational efficiency. Those uncertainties are introduced because for dense systems typically only a very limited number of moves are used, which are sometimes limited to only an interchange of two vertical parallel bonds by two horizontal ones and vice versa. This limitation makes the techniques slowly converging, since the limited collection of Monte Carlo moves builds a long correlation between the Monte Carlo steps, and, can also possibly result in error, since ergodicity is still uncertain[16]. Furthermore, no absolute thermodynamic potentials, such as entropy, can be calculated from the application of such a Monte Carlo approach alone. The capability to evaluate the thermodynamic quantities is very necessary if these models are to be applied to study structure-property relationships within a non-equilibrium thermodynamic framework[21-22].

Earlier, we had developed an alternative computational approach to study dense semicrystalline polymer conformations using lattice models[4]. It was based upon an exhaustive enumeration of all possible microscopic configurations of the polymer chains, which can be performed efficiently for small two-dimensional lattices in a computer by an algorithm that treats exactly the chain connectivity and the excluded volume constraints for fully populated lattices. In this approach, information is generated in terms of the permutations of the vertical and the horizontal bonds in the lattice rows (i.e. in terms of the polymer segments layout in the lattice) instead of in terms of tracing individual chains. A major limitation of the above approach is that each macroscopic state is microscopically specified very tightly, by fixing individually the number of vertical bonds connecting each lattice row to the one beneath. As a consequence, thermodynamic quantities, such as entropy and free energy, calculated subject to this constraint, may require unreasonably large lattice sizes to converge to the thermodynamic limit. Therefore, in this work, we present a much more efficient algorithm based on a lattice subdivision technique. This technique still allows for an exact evaluation of thermodynamic quantities for small lattices through a systematic generation of all possible chain conformations. When no energetic effects are considered, a stochastic enumeration approach has also been developed that allows us to use larger lattices. Moreover, for studying larger lattices even in the presence of non-zero energetics, we have developed a Metropolis-based Monte Carlo[23] scheme. An advantage of the exhaustive enumeration method is that it allows for the validation of both the stochastic enumeration and the Metropolis-based Monte Carlo method at least for small lattice sizes.

In this paper we present a two dimensional (2-D) implementation of our approach. We have performed a rigorous validation of the Monte Carlo scheme against exact enumeration and stochastic enumeration approaches. We also offer quantitative analysis of the Monte Carlo results based on a rigorous evaluation of the correlation between the Monte Carlo realizations. This allows for simultaneous determination of tight error bars as various mean statistics are calculated. In the second paper[24] of this series we apply this approach to study the lattice size and energy effects on 2-D lattice simulations. There, we obtain statistics for chain conformations and thermodynamic potentials associated with the interlamellar amorphous region of semicrystalline polymers for various lateral lattice sizes thus leading to predictions in the thermodynamic limit (corresponding to an infinite lateral dimension). These results are extended to three-dimensional lattices in the third paper[25] of the sequence.

We begin by describing the mathematical problem in section 2. The solution methodology is outlined in section 3. Typical results obtained with the proposed approach are presented in section 4. The conclusions are presented in section 5.

## 2 Problem Definition and Terminology

We consider, for simplicity, a square lattice of size $L$x$N_0$, with $L$ rows each with $N_0$ nodes long sandwiched between two crystallites, as shown in Figure 1. Rows 1 and $L$ are considered adjacent to crystalline lattices and have vertical bonds emanating downwards and upwards respectively. Other crystalline chain orientations can also easily be studied, if so wished, simply by suitably modifying the boundary conditions; for example, alternating horizontal and vertical bonds represent chains inclined at $45^0$.

The chains originating from the crystalline interface are assumed to continue in the amorphous region represented by the bulk of the lattice. The lattice is completely and regularly filled, i.e. each lattice site must be connected to two and only two (out of four) neighboring lattice sites, and as such no density variations between the amorphous and the crystalline regions are allowed. Introduction of chain ends that can possibly lead to density variations is left for future investigations. The chains are thus assumed to be long enough so that there are no free ends within the lattice. As commonly practiced, periodic boundary conditions (BCs) are imposed along the other two (vertical) boundaries of the lattice in order to simulate a lamella infinite in the lateral direction. The periodicity introduces a tighter constraint, which needs to be relaxed (for example, by taking the limit $N_0 \rightarrow \infty$) in order to obtain results completely free of finite size effects.

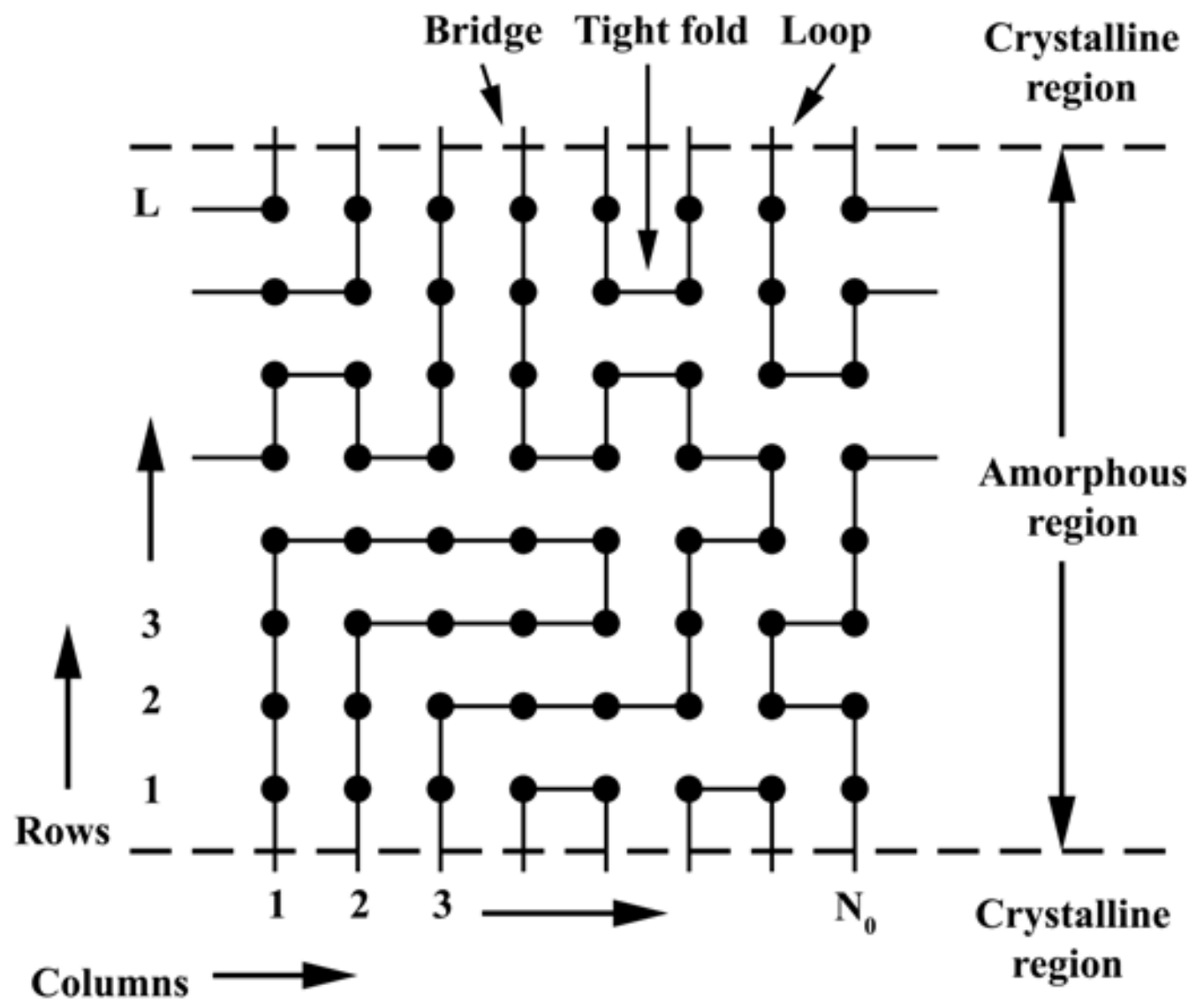

**Figure 1**. A sample microstate corresponding to a 2-D interlamellar lattice of size $L$x$N_0$, with the amorphous phase $L$ (=8 here) rows deep with periodicity enforced in the horizontal direction. There are $N_0$ (=8 here) lattice sites in a row. Note that the lattice is fully populated and that the connectivity requirements at the crystalline boundaries are also satisfied.

A connection between two adjacent lattice sites in the same row will be referred to as a *horizontal bond* whereas a connection between two adjacent lattice sites in the same column will be referred to as a *vertical bond*. We refer to a feasible arrangement of vertical and horizontal bonds, which satisfies the space-filling constraints mentioned above and is devoid of closed loops (rings), as a *microstate*. One such microstate is shown in Figure 1. In any particular microstate, a chain connecting the two crystalline boundaries is said to form a *bridge* whereas a chain starting from one crystalline boundary and returning to the same crystalline boundary is said to form a *loop*.

On the other hand, despite the fact that the chain segments can take any one of the allowed orientations by the coordination of the lattice, intra-molecular strain effects can also be modeled by using energy penalties that are specific to individual chain conformations. More specifically, in this work we follow the established practice to use an energy penalty proportional to the number of "tight folds" in the chain conformations. We define as a tight fold a bond in the middle of a 3-bond sequence in form of a “Π” as shown in Figure 1. These energetic terms affect the Boltzmann factors that are used to penalize the corresponding conformation probabilities.

Our aim is to find the average chain statistics, corresponding to given amorphous-crystalline boundary conditions and energy conditions. To do so we propose a new computational methodology based on a subdivision of the original 2-D lattice into smaller equal size sublattices. The calculation of the microstates for the entire lattice is then performed based on the information for the submicrostates (i.e. microstates for all possible sublattices), which is generated separately and stored in databases. This algorithmic approach is described in more detail in the next section.

## 3 Solution Methodology

### 3.1 Lattice subdivision

First, the lattice *($L$x$N_0$)* is divided into smaller sublattices as shown in Figure 2. The size of a sublattice has been chosen here to be 4x4 as the best compromise between a large enough size to make the construction of lattice microstates computationally feasible and a small enough one to result in databases of manageable size. For simplicity, in the current application of the algorithm we also require that both $L$ and $N_0$ be divisible by 4. A slightly more involved algorithm using specialized databases can also be easily established, if needed, to circumvent this limitation.

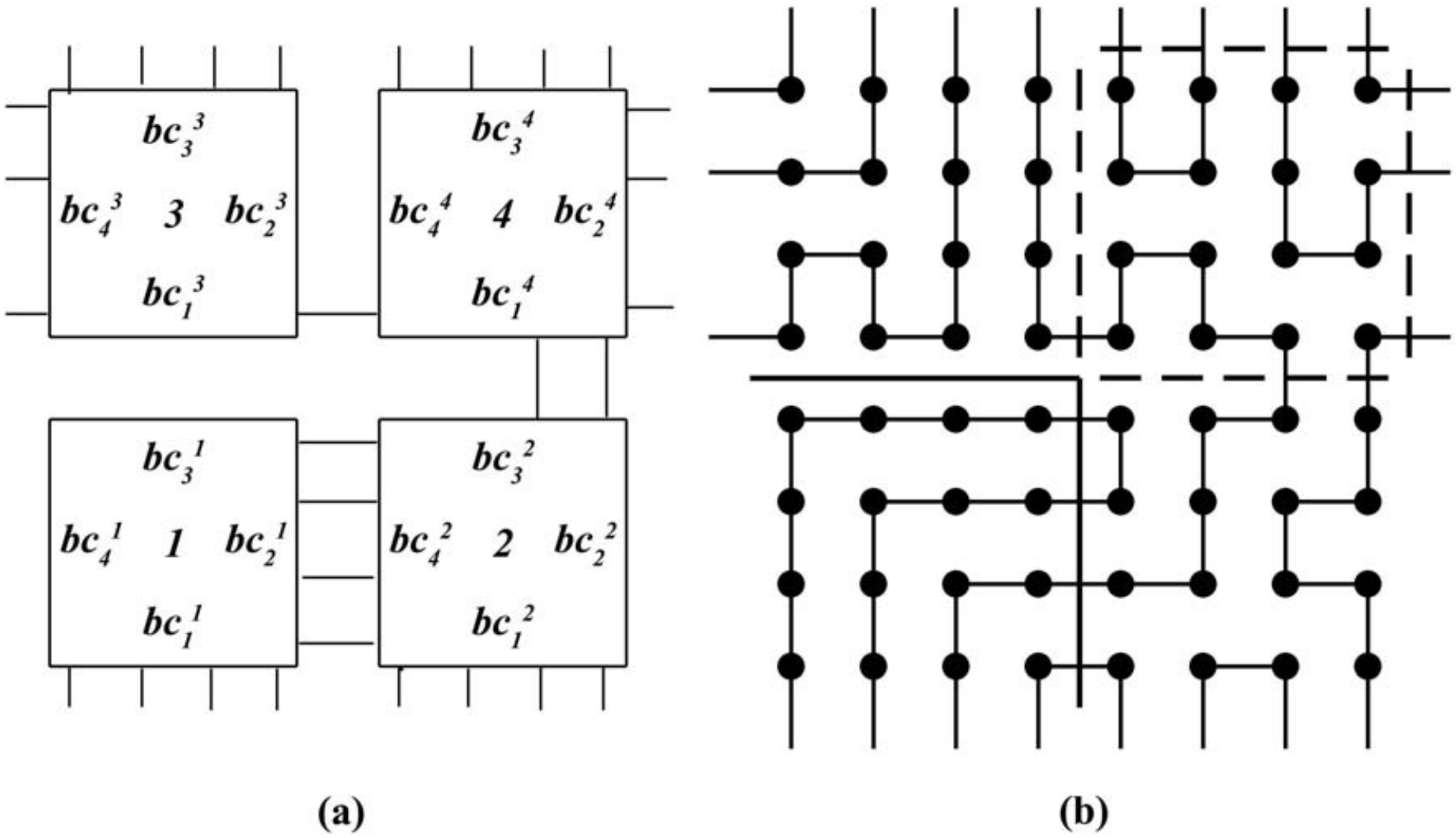

**Figure 2.** A schematic representation of the lattice subdivision into sublattices applied to a particular bond realization. (a) sublattice connectivity and notation (b) Detailed bond information for one possible internal realization.

A microstate for the entire lattice can be generated by choosing those *submicrostates* (microstate of a 4x4 node sublattice) for the individual sublattices that "connect" consistently with each other and give rise to a system of chain conformation that are devoid of rings. A given set of inter-sublattice connections defines the boundary conditions (BCs) necessary for the generation of the submicrostates on that sublattice. Here, a boundary condition is a permutation of a sequence of "holes" (i.e. corresponding to an absence of connecting bonds) and normal bonds emanating perpendicularly from the boundary of the sublattice under consideration. Proceeding in a counterclockwise fashion starting from the bottom we denote each permutation of boundary conditions (normal bonds and holes) surrounding sublattice $j$ as $bc_1^j$, $bc_2^j$, $bc_3^j$, $bc_4^j$, see Figure 2(a).

### 3.2 Database generation of the submicrostates of a 4x4 sublattice

Information on all possible submicrostates of the 4x4 sublattice compatible to all possible BCs is generated beforehand and stored in look-up tables. This significantly increases the computational efficiency of the simulation for the entire lattice, which simply involves a selection of submicrostates for the sublattices from this database. The side BCs of the sublattice is uniquely specified by a single hexadecimal digit, i.e. an integer number in the range [0,15], which is constructed when the corresponding permutation of the occupied normal bonds emanating from that boundary is represented by a four digit binary number with 1 representing the presence of a normal bond at a particular position and 0 an absence. Then, for each particular set of the four BCs of a sublattice, represented collectively by a four digit hexadecimal number, $\{bc_1,bc_2,bc_3,bc_4\}$, corresponding to the four sublattice edges, all feasible submicrostates of the 4x4 sublattice are generated and stored. The database is organized in such a way that it provides information about the total number of submicrostates and the layout of these submicrostates for any set of desired BCs. Additional information needed for the exact enumeration or the Monte Carlo approach, which is pertinent to individual sublattice microstructures (such as the connectivity between the entering-exiting boundary bonds corresponding to the same polymer chain), is also generated *a priori* and stored in separate databases. Finally, note that at this stage we also check and discard all internal bond arrangements containing rings.

### 3.3 Constructions and counting of the microstates of the lattice

*3.3.1 Exhaustive enumeration*

For small lattice sizes we can achieve an exhaustive enumeration of all microstates, i.e. we can generate all possible ways of filling up the lattice. This is performed, using lattice subdivision technique, by allowing for all possible interconnections (represented by different boundary conditions, BCs) and subsequently, by selecting for each set of BCs all possible submicrostates from the previously constructed databases. This process is subsequently followed by a check for rings, which may form as a result of connecting two or more sublattices, and discarding those lattice conformations that contain rings.

The compatibility in the connections along a shared boundary between neighboring sublattices results in the requirement that the same BCs apply for adjacent sublattice edges (Figure 2). In addition some BCs of a sublattice are fixed, viz. those at the amorphous-crystal boundary while laterally at the far left and far right of the computational domain we apply the periodic boundary conditions. For example in Figure 2 (a), which shows a division of an 8x8 node lattice into 4 sublattices of 4x4 nodes, $bc_1^1$, $bc_1^2$, $bc_3^3$, $bc_3^4$ are fixed by the amorphous-crystal interface, $bc_1^3= bc_3^1$, $bc_4^2= bc_2^1$, $bc_4^4= bc_2^3$, $bc_1^4= bc_3^2$ are imposed to satisfy the connectivity requirements and $bc_2^2= bc_4^1$, $bc_2^4= bc_4^3$ due to the periodicity conditions in the lateral direction. Once a set of boundary conditions is selected compatible with the abovementioned constraints, the selection of lattice microstates is done by sequentially selecting submicrostates (microstate of a sublattice) from previously generated databases. In the snapshot shown in Figure 2 (b), one possible set of specific sublattice microstates compatible to the boundary conditions shown in Figure 2(a) is shown. One of the sublattices (sublattice 4) is especially emphasized with a dashed box. Note that the particular selection of the submicrostates shown in Figure 2(b) is an allowed one as it does not contain any chain rings-a check for rings always follows the submicrostate selection as rings are disallowed.

The chain statistics are then collected for each allowed lattice microstate and appropriately averaged. Since in the current work only tight fold energetics are taken into account, the microstates are catalogued with respect to the total number of tight folds present in them. The statistics are then calculated in a post-processing step by weighing each average corresponding to $N_{folds}$ number of tight folds by the corresponding Boltzmann probability, $exp(-E_\eta N_{folds})$, where, $E_\eta$ is the energetic penalty per tight fold non-dimensionalized with respect to the Boltzmann factor, $k_BT$, and $N_{folds}$ is the total number of tight folds in the given microstructure, thus allowing us to calculate statistics for arbitrary $E_\eta$ without having to re-evaluate all possible lattice microstates.

The advantage of the exhaustive enumeration is that it yields exact results and therefore forms a yardstick against which other approximate techniques can be compared. Furthermore, the widespread use of databases makes the scheme very efficient. Even so, the exponentially fast increasing number of microstates with increasing lattice size severely limits the lattice size where this exact enumeration approach can be applied, no matter how efficient each microstate calculation is. In the particular examples considered here, i.e. for the amorphous region in a 2-D interlamellar lattice, this limitation translates to lattice sizes of 8x8 nodes. The results from exhaustive enumeration for 8x8 node lattice are used in this work to validate the stochastic enumeration and Monte Carlo techniques, which are described next.

*3.3.2 Stochastic enumeration*

As mentioned before, the exponential increase in computational time with lattice size to get exact estimates necessitates a different calculating approach to allow larger lattice sizes to be investigated. This is met in this work through the development of a stochastic enumeration, discussed here, and a Monte Carlo technique, described in the following section. The stochastic enumeration is based on an unbiased sampling technique. The basis of the algorithm can be described by the following steps, as they are applied to evaluate stochastically the microstates in a lattice containing $L_s$x$N_s$ sublattices corresponding to a height of $L = 4L_s$ and to a lateral dimension of $N_0 = 4N_s$ nodes:

1. For each sublattice $k$; $k=1, L_s$x$N_s$

   We first estimate $w_k$, the number of submicrostates that satisfy already selected BCs in the following manner:

   1.1 Using information from databases we find all possible permutations of BCs (say $N_k$) for sublattice $k$, that are compatible with any already selected ones. We randomly select one of the $N_k$ permutations and assign it to the sublattice $k$. This step automatically fixes in the end all boundary conditions of sublattice $k$.

   1.2 Using information from databases, we find the number of submicrostates compatible with the BC's of this sublattice (say $M_k$). We randomly select one of the $M_k$ permutations and fill up subblattice $k$.

   1.3 We evaluate the total weighting factor corresponding to that sublattice microstate, $w_k$ as, $w_k = N_k M_k$

2. After all sublattices have been considered, the total weighting factor $W_i = \prod_{k=1}^{N_s \times L_s} w_k$ of all submicrostates that have been selected is calculated. $W_i$ may be zero if at any point in the process no internal states are found compatible to the existing set of BCs. If $W_i = 0$, we set $r_i = 0$ and we go to step 4, else we proceed to step 3.

3. Based on all submicrostates internal chain information we check for rings. If the lattice microstate is devoid of rings, then microstate statistics are collected and $r_i$ is set equal to1. Else, the case is disregard and $r_i$ is set equal to 0.
4. Steps 1-3, are repeated for $i = 1, 2, 3, \ldots$ a predetermined $N_{sample}$ number of times.

However, the underlying assumption here is that *a priori* all microstates are equally probable. This is true only in the absence of energetics. Thus the stochastic enumeration technique can only be applied when no energetic effects are present (i.e. in the purely entropic limit). Even so, the technique is useful because it can be used to estimate the absolute conformational entropy, $S$, of systems without energetic interactions. This is defined as usual in terms of the total number of microstates, $\Omega$, as $S = k_B \ln \Omega$ (see section 4.1 for an example). This, in turn, can be used, in conjunction with the Monte Carlo scheme discussed below, to determine the total free energy even in the presence of energetic effects. An important issue in any stochastic simulation scheme is to get reliable estimates of the error bounds associated with the predictions. This is discussed in the next section.

*3.3.3 Estimation of the standard deviation associated with stochastic enumeration*

In the stochastic enumeration scheme presented above there is a weight $W_i$ associated with every set of independent boundary conditions, equal to the number of possible microstates compatible to those conditions. Considering further a random variable '$r$' which takes a value of 1 if there are no rings for the particular lattice microstate and 0 otherwise, leads to an unbiased estimator for the population size, $N_{est}$ given by

$$N_{est} = \frac{\sum_{i=1}^{N_{sample}} W_i r_i}{N_{sample}} , \qquad \ldots 1$$

with the corresponding variance given by

$$\sigma_{N_{est}}^{\ 2} = \frac{\dfrac{\sum_{i=1}^{N_{sample}} \left(W_i r_i\right)^2}{N_{sample}} - N_{est}^{\ 2}}{N_{sample} - 1} . \qquad \ldots 2$$

Here $r_i$ is an independent random variable obtained with a probability proportional to1/ $W_i$. This follows standard statistics. However, care needs to be exercised when the standard deviation of various chain statistics is determined since those represent in stochastic enumeration approach, composite quantities (i.e. obtained by taking ratio of averages) since the random variables corresponding to these averages are not independent.

For illustration purposes, assume that random variable $x$ represents the weight associated to any random sampling, $x_i = W_i r_i$, and that variable $y$ represents the corresponding weighted statistics, $y_i = W_i r_i s_i$, *where* $s_i$ is any microstate property. Then a corresponding unbiased estimator of s is given by $\mu_{y,x}$ which is defined as,

$$\mu_{y,x} = \frac{\dfrac{\sum_{i=1}^{N_{sample}} y_i}{N_{sample}}}{\dfrac{\sum_{i=1}^{N_{sample}} x_i}{N_{sample}}} = \frac{\sum_{i=1}^{N_{sample}} y_i}{\sum_{i=1}^{N_{sample}} x_i} \quad . \qquad \ldots 3$$

Let $\bar{x} = \langle x \rangle$ and $\bar{y} = \langle y \rangle$, and *x'* and *y',* denote the average and the fluctuating (deviation from the mean) quantities for *u* and *v,* respectively. Then *u'* can be linearly decomposed into two random variables as:

$$y' = \frac{\sigma_y}{\sigma_x}\left(cx' + \sqrt{1-c^2}\,z'\right) \quad , \qquad \ldots 4$$

where $\sigma_x$ and $\sigma_y$ are the standard deviations of the respective variables, *c* is the correlation coefficient between *x* and *y,* $c = \frac{\langle x'y' \rangle}{\sigma_x \sigma_y} = \frac{\langle xy \rangle - \langle x \rangle\langle y \rangle}{\sigma_x \sigma_y} = \frac{\mathrm{Cov}(x,y)}{\sigma_x \sigma_y}$*,* and *z* is another independent random variable defined by this formal decomposition.

Based on the definition of $\sigma_x, \sigma_y$ and *c* it follows that

$\langle z'x' \rangle = 0$; $\langle z' \rangle = 0$; $\langle z^2 \rangle = \sigma_x^2$, i.e. *z* and *x* are independent random variables.

Then, in the limit of $\frac{x'}{\bar{x}} \ll 1$, equations 3 and 4 can be used to evaluate an approximate expression for the variance of $y/x$ as:

$$\sigma_{y/x}^2 = \frac{\bar{y}^2\sigma_x^2 + \bar{x}^2\sigma_y^2 - 2c\overline{xy}\sigma_x\sigma_y + \left(1-c^2\right)\sigma_x^2\sigma_y^2}{\bar{x}^4} \quad , \qquad \ldots 5$$

which can be further simplified to

$$\sigma_{y/x}^2 = \frac{\langle x^2 \rangle\langle y^2 \rangle - \langle xy \rangle^2}{\bar{x}^4} \quad . \qquad \ldots 6$$

However, we are interested in obtaining an expression for the variance of $\mu_{y,x}$, which is defined in equation (3) as the ratio of two mean quantities rather than the ratio of individual random realizations. In this case, the variance of $\mu_{y,x}$, $\sigma_{\mu_{y,x}}^2$, can be determined (for details see appendix B of ref [25]) for large $N_{sample}$ as:

$$\sigma_{\mu_{y,x}}^2 = \frac{\bar{y}^2\sigma_x^2 + \bar{x}^2\sigma_y^2 - 2c\overline{xy}\sigma_x\sigma_y}{N_{sample}\bar{x}^4} \quad . \qquad \ldots 7$$

In all cases that we tested it, equation 7 has proven to provide an accurate a priori estimate of error associated with the statistics calculated by the stochastic enumeration technique- see also results of validation studies in section 4.1.

*3.3.4 Monte Carlo scheme*

The key elements to the success of any Monte Carlo scheme are its ergodicity and computational efficiency. The latter is dependent on the computational load in generating a move and the correlation between subsequent moves. In general, the easier a move is generated, the higher is the probability that the scheme is not ergodic, resulting in a biased answer, and the higher is the correlation between subsequent moves, resulting in a larger variance. The proposed set of moves has therefore been devised in order to strike the best compromise between low correlations between moves (with a high probability for the scheme to be ergodic) and easy move generation.

The following classes of Monte Carlo moves are thus proposed:

i. Sub-lattice optimization: This involves first selecting randomly a lattice subdivision into sublattices, by shifting the lattice horizontally and/or vertically by a randomly selected number of columns and/or rows, respectively, followed by a regular subdivision of the transformed lattice to 4x4 node sublattices starting from the low left corner. A new microstate is then formed in either one of two possible ways: a) through a global selection by choosing randomly for each one of the sublattices a submicrostate that conforms to the fixed connectivity requirements (boundary conditions) corresponding to the previous state or b) by randomly selecting one of the sublattices, then only selecting a new submicrostate in that sublattice among all submicrostates conformed to that sublattice's BCs, keeping the rest of the lattice, including the connectivity, the same. At the end of either (a) or (b) step, the newly proposed global microstate, if different from the previous one, is checked against possible rings between sublattices (since all submicrostates are devoid of internal rings by construction). If a ring is found, any changes made are annulled and the submicrostate selection procedure repeats itself. Note that there is always convergence to this iteration, since there is at least one microstate, the initial state, as an accepted solution. Once an acceptable solution is chosen, an "unshift" operation is performed to recover the original lattice layout of rows and columns. Thus, the purpose of the shift and unshift operations is simply to allow for a richer lattice subdivision into sublattices, making sure that many chances are provided for all bonds to change.

ii. Symmetry transformation: This involves consideration of various possible symmetry operations in sequence, such as a mirror imaging (left-right and/or up-down) of the lattice structure and/or a permanent shifting of the microstructure by a randomly chosen number of columns exploiting the periodicity in the lateral direction. The symmetry operation does not produce a new geometrical configuration and as such does not enhance the ergodicity of the scheme, and hence it is not absolutely necessary. However, it does allow switching effectively between all notationally different but geometrically equivalent configurations, leading to more symmetric results for the same number of Monte Carlo steps.

iii. Isomerization: This step involves replacing any combination of rows that possibly exist involving a sequence of alternating void or filled horizontal bonds with their complements where the alternation is exchanged. The purpose of that move is to blend populations that otherwise will have been disjoint if only the other two types of moves were to be used.

To further illustrate these moves and their effectiveness in sampling the entire population of microstates we consider here an implementation of this scheme on a toy 2 x 4 node lattice subdivided into two 2 x 2 node lattice. The seven possible submicrostates on a 2x2 sublattice with crystalline top and bottom boundary conditions are shown in Figure 3. The seven sublattices can be combined to form nine lattice microstates for the 2x4 node system, which are shown in Figure 4. A set of Monte Carlo moves sublattice optimization in all or one sublattice, symmetry operation, isomerization was implemented. It was found that two of the nine lattice microstates (states B and C) form a separate subpopulation in the absence of isomerization move. However, when the isomerization move is used, all nine states can be reached regardless of the initial guess as a perusal of the resulting transition space indicates (Table 1).

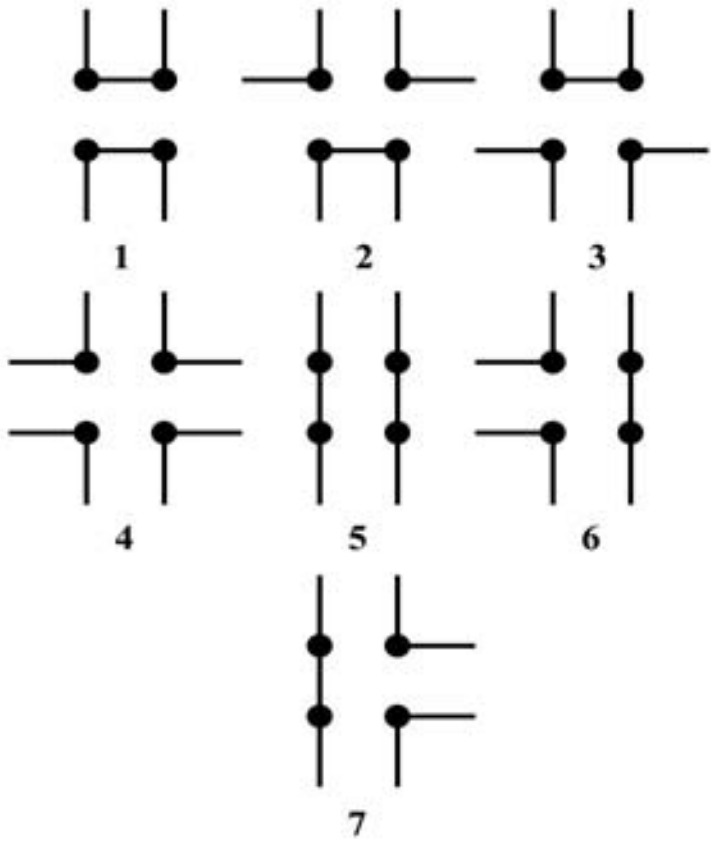

**Figure 3.** The seven submicrostates for 2x2 node sublattice compatible to top and bottom crystalline BCs.

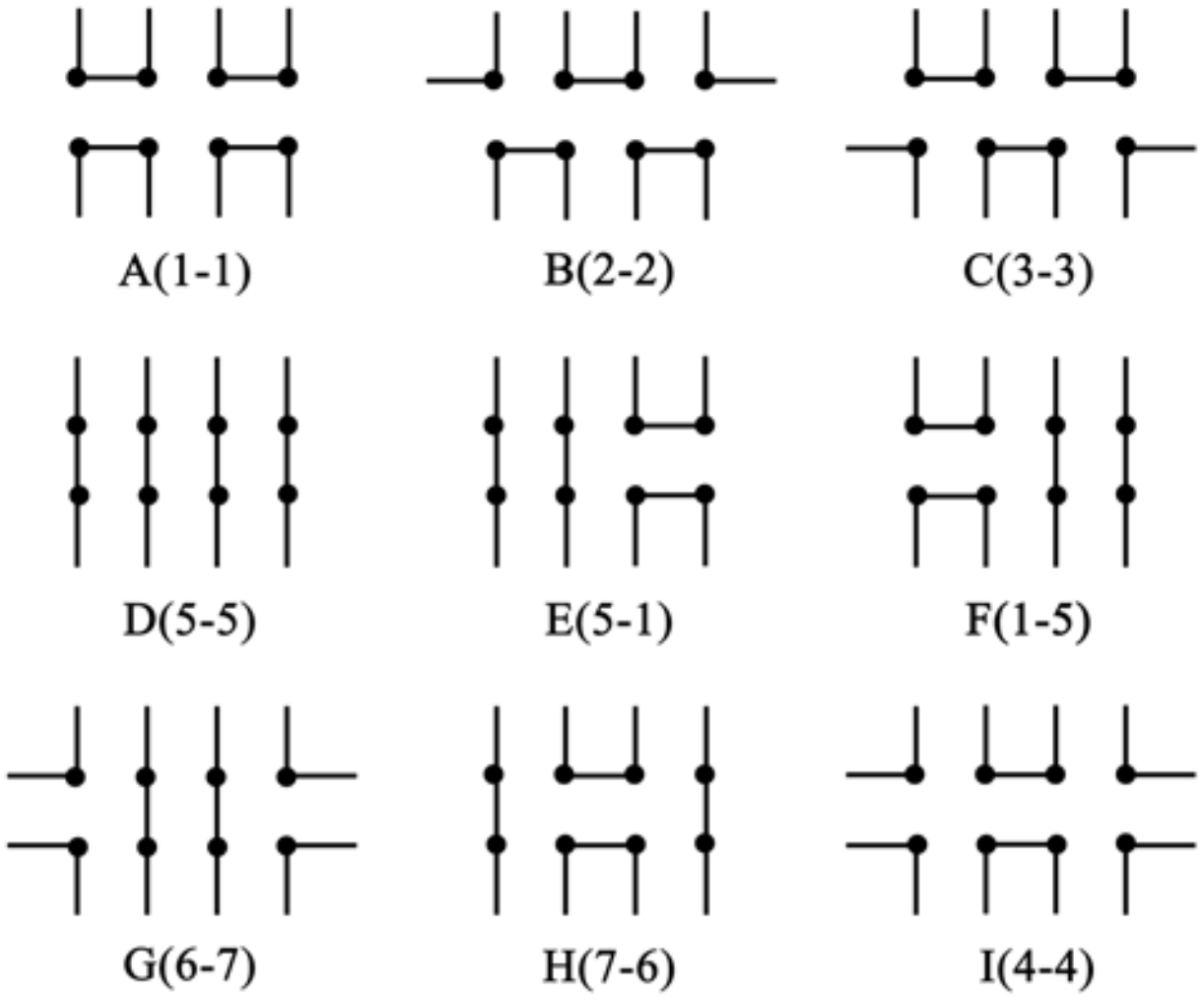

**Figure 4.** The nine microstates for 2x4 node lattice constructed from the seven possible 2x2 node sublattice.

**Table 1.** Implementation of Monte Carlo scheme in absence and in presence of the isomerization move. Note that states B and C form an isolated subpopulation in absence of isomerization move.

| Starting state | Sublattice optimization in all sublattices | Sublattice optimization in one sublattice | Symmetry transformation | Isomerization |
|---|---|---|---|---|
| A | A,E,F,D | A,E,F | A,I | A,B,C,I |
| B | B | B | B,C | B,A,C,I |
| C | C | C | B,C | C,A,B,I |
| D | D,A,E,F,G,H,I | D,E,F,G,H | D | D |
| E | E,D,A,F | E,D,A | E,F,G,H | E |
| F | F,E,D,A | F,D,A | F,E,G,H | F |
| G | G,H,I,D | G,H,I | G,E,F,H | G |
| H | H,G,I,D | H,G,I | H,E,F,G | H |
| I | I,G,H,D | I,G,H | I,A | I,B,C,A |

**Table 2.**Transition probability matrix for the Monte Carlo scheme implemented on 2x4 node lattice for $p_{all}$ = 0.4, $p_1$ = 0, $p_s$ = 0.2, $p_{iso}$ = 0.4.

| | A | B | C | D | E | F | G | H | I |
|---|---|---|---|---|---|---|---|---|---|
| A | 0.45 | 0.10 | 0.10 | 0.05 | 0.05 | 0.05 | 0.00 | 0.00 | 0.20 |
| B | 0.10 | 0.60 | 0.20 | 0.00 | 0.00 | 0.00 | 0.00 | 0.00 | 0.10 |
| C | 0.10 | 0.20 | 0.60 | 0.00 | 0.00 | 0.00 | 0.00 | 0.00 | 0.10 |
| D | 0.05 | 0.00 | 0.00 | 0.70 | 0.05 | 0.05 | 0.05 | 0.05 | 0.05 |
| E | 0.05 | 0.00 | 0.00 | 0.05 | 0.70 | 0.10 | 0.05 | 0.05 | 0.00 |
| F | 0.05 | 0.00 | 0.00 | 0.05 | 0.10 | 0.70 | 0.05 | 0.05 | 0.00 |
| G | 0.00 | 0.00 | 0.00 | 0.05 | 0.05 | 0.05 | 0.70 | 0.10 | 0.05 |
| H | 0.00 | 0.00 | 0.00 | 0.05 | 0.05 | 0.05 | 0.10 | 0.70 | 0.05 |
| I | 0.20 | 0.10 | 0.10 | 0.05 | 0.00 | 0.00 | 0.05 | 0.05 | 0.45 |

In Table 2 we present the full transition probability matrix obtained for the 2x4 node lattice system obtained with $p_{all,}$ the probability with which sublattice optimization is performed in all sublattices (2 here), being 0.4, $p_1$, the probability of performing sublattice optimization in one sublattice, equal to 0 and $p_s$, the probability of performing symmetry transformation being 0.2, thus corresponding to $p_{iso}$, the probability with which isomerization is performed, equal to 0.4. Any three independent parameters, like $p_{all}$, $p_1$, $p_s$, can be used to fully characterize the Monte Carlo method, the fourth being evaluated from the normalization criterion: $p_{iso} = 1 - p_{all} - p_1 - p_s$ in our case. The matrix derived for the non-energetic case, is symmetric and provides full connectivity, thus generating unbiased results from any initial guess.

Incorporated to anyone of the above mentioned moves for which there is a possibility for a geometrically distinguished microstate (i.e. in moves i and iii), is a selection bias procedure following the Metropolis algorithm[23] with a microstructure probability proportional to $exp(-E_\eta N_{folds})$, where $E_\eta$ is the energetic penalty per tight fold normalized with respect to the Boltzmann factor, $k_BT$, and $N_{folds}$ is the total number of tight folds in the microstructure. In particular, associated with the sublattice optimization step with option (b) we use a variant of the traditional Metropolis algorithm according to which at each Monte Carlo step we allow for a weighted selection among all of the connectivity compatible submicrostates. More specifically all of the allowed configurations within a randomly selected sublattice are allowed to be chosen with probabilities pre-weighted by their corresponding Boltzmann factor, thus allowing for a faster transition to the statistical equilibrium distribution.

As explained above for an ergodic scheme isomerization move probability $p_{iso}$ needs to be non-zero. Nevertheless, $p_{iso}$ is chosen small, because it rarely leads to a successful change, and hence it does not disturb the effectiveness of the sublattice isomerization steps. Another important issue that has been missing in all previous Monte Carlo schemes is the accurate estimation of the error associated with the sampling process. This is very important in order to be able to extract quantitatively reliable predictions and it is discussed in the next section.

*3.3.5 Estimation of standard deviation associated with the Monte Carlo scheme*

In order to obtain quantitatively reliable estimates of any quantity in Monte Carlo simulations, it is first necessary to evaluate the error associated with the averages. One way to estimate the error would be an aposteriori estimation based on results of several independent simulations. This method is obviously quite time consuming. Here we discuss the way to evaluate the errors simultaneously with the Monte Carlo simulation.

The average of any statistical quantity, $x$ obtained using an $N_{sample}$ strong Monte Carlo sampling is easily calculated using the unbiased estimator, $\mu_{x,N_{sample}}$ given by

$$\mu_{x,N_{sample}} = \frac{\sum_{i=1}^{N_{sample}} x_i}{N_{sample}} \quad . \qquad \ldots 8$$

However the evaluation of the standard deviation associated with that estimator, $\sigma_{\mu_{x,N_{sample}}}$, is more involved as there exists, in general, non-zero correlations between subsequent Monte Carlo moves. Taking these into account, the variance of the statistics in Monte Carlo sampling is given by the following equation where $c_i$ is the pair correlation coefficient between samples separated by $i$ number of Monte Carlo moves given by equations 8 and 9.

$$\sigma^2_{\mu_{x,N_{sample}}} = \frac{1}{N_{sample}-1}\left(<x^2> - \mu_{x,N_{sample}}{}^2\right)\left(1+2c_1+2c_2+2c_3+...2c_{N_{sample}-1}\right) \quad , \qquad \ldots 9$$

$$c_i = \frac{\left(<x_j x_{j+i}> - <x>^2\right)}{<x^2> - <x>^2} \quad , \qquad \ldots 10$$

$$<x^2> = \frac{1}{N_{sample}} \sum_{j=1}^{N_{sample}} x_j{}^2 \quad . \qquad \ldots 11$$

The above expressions can also be found in standard statistics textbooks- see for example page 67 of ref. [26].

## 4. Validation Results

Given the availability of the exact results for small enough lattice sizes following the exhaustive enumeration procedure, we report here primarily the results of a systematic validation procedure of the stochastic enumeration and the Monte Carlo procedures. In particular, in subsections 4.1 and 4.2 we test quantitatively the results of the stochastic approaches developed in this work not only with respect to the estimates of the averages they provide but also with respect to independently developed measures for the accompanying error. This is very important if quantitative results are to be extracted in large lattice Monte Carlo runs. In part –2 of this series[24] we present results pertaining to the evaluation of quantities in the thermodynamic limit (in the limit of infinite lateral extent of the lattice), which is the main motivation behind this work and illustrates the usefulness, and feasibility of the developed approach.

### 4.1 Validation of the stochastic enumeration technique

A comparison between the exact results obtained with the exhaustive enumeration technique and the approximate one using the stochastic enumeration is shown in Figure 5 and 6 for an 8x8 node lattice as far as a state statistics (average number of vertical bonds from below in a row) and the total population size respectively are concerned.

As can be seen from Figure 5 and 6, there is an excellent agreement and full consistency in the stochastic enumeration results. Not only the statistical averages are close to the exact answers, but also the latter are always seen to lie within the independently determined error bars irrespective of the seed to the random number generator and sample size. Furthermore, the error decreases consistently by a factor of 2 as the sampling size is increased by a factor of 4. It is also important to note here that the error bars (taken at the 95% confidence level, i.e. within two standard deviations) are calculated based on the results of the stochastic enumeration, independently of the exact results. Thus, the fact that in all cases, the exact results are found to span more or less the full error margin, provides a strong evidence for the unbiased character of the sampling (which should be the case according to the theory of stochastic enumeration procedure for the non-energetic interactions studied here) and validates the approximations used in the error estimation.

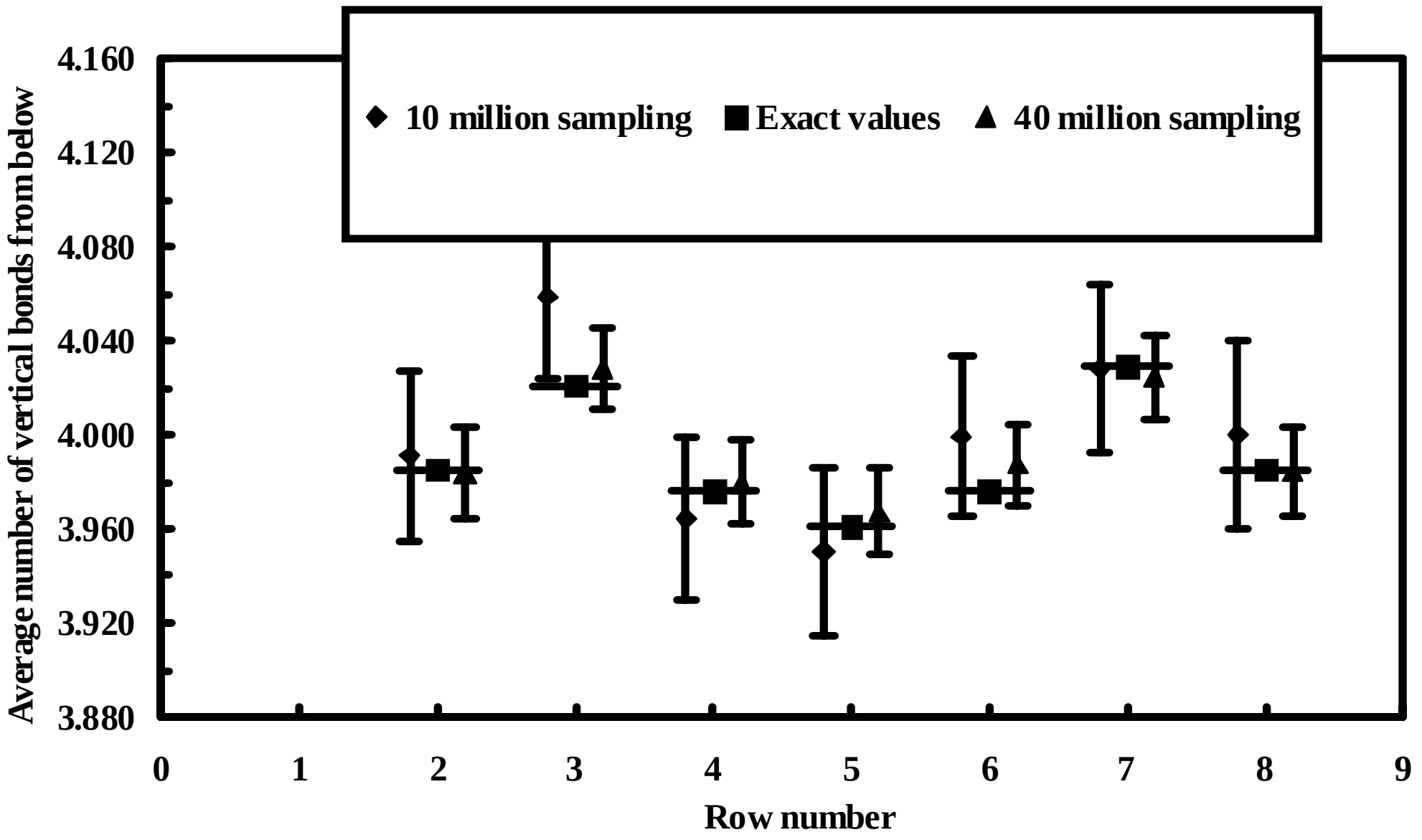

**Figure 5.** Comparison of the predictions obtained with two stochastic runs using 10 and 40 million samples against the exact results for the average number of vertical bonds connecting a row from below, for each row of an 8x8 node amorphous lattice. The stochastic results are reported within error bars reported here at ± two times the standard deviation The standard deviation is approximated stochastically simultaneously with the reported approximations for the mean.

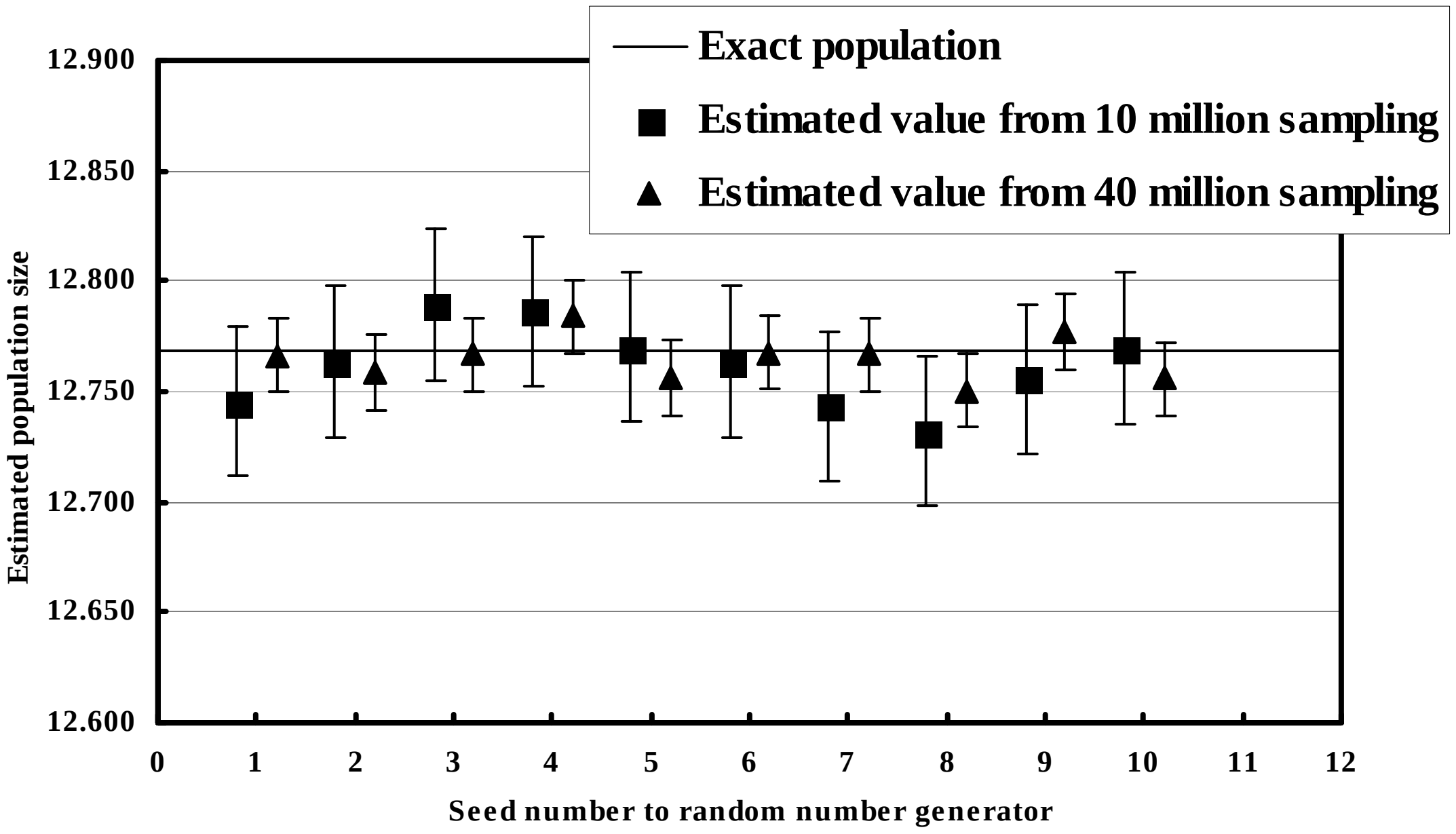

**Figure 6.** Comparison of the predictions obtained with two series of stochastic runs using 10 and 40 million samples against the exact results for the total number of microstates for an 8x8 node lattice. The predictions are described with respect to the seed number of the random number generator used to start the sampling process and are reported with error bars approximated as ± two times the standard deviation, also determined stochastically.

## 4.2 Validation of the Monte Carlo scheme

An essential element for any Monte Carlo scheme is whether the scheme is ergodic. Unfortunately this is very difficult, if not impossible, to prove in the general case. Nevertheless, evidence can be collected for special cases. In our work we did that by applying our Monte Carlo scheme to small enough lattices where the exhaustive enumeration procedure could be applied to estimate exact answers by counting all possible microstates of the system. In particular two lattice sizes were studied in detail, 4x8 and 8x8 as discussed in the following subsections

### *4.2.1 4x8 node lattice investigations*

From the exact results for this system using the exhaustive enumeration algorithm, this system was found to have exactly 34285 microstates. The Monte Carlo scheme was then carried out for a number of steps much higher than the population size (7500 times) and the frequencies with which each microstate is visited in the Monte Carlo scheme were recorded. Since no energetic effects were considered, all microstates were equally probable and any ergodic Monte Carlo scheme must have sampled each of them equally. Indeed the relative frequency (relative to the expected value 7500) with which any microstate was visited was seen to vary between 0.935 and 1.065 averaging at 1.0, demonstrating the ergodicity of our scheme. Furthermore the distribution of the microstates with respect to the relative frequency was almost a Gaussian, with a peak at 1.0 and a spread of around 0.07. Also the spread in the distribution indicated a correlation length of 5, which matches well with the independently evaluated correlation length of 4 steps. Finally, the predictions of the Monte Carlo scheme were compared with the exact values obtained using the exhaustive enumeration scheme. The Monte Carlo predictions were close to the exact values with error bars always capturing the exact answer. More such validation results in the case of predictions for an 8x8 node lattice are discussed next.

### *4.2.2 8x8 node lattice investigations*

An 8x8 node lattice is one of the largest sizes that can be investigated exactly. Even for such a relatively small lattice size there are 12,767,878,957 microstates. The same lattice was analyzed using the Monte Carlo scheme. For the runs reported here, the parameters for the Monte Carlo move selection, $p_{all}$, $p_1$, $p_s$ were set to 0.85,0.0,and 0.10, respectively, with the rest (i.e. 0.05), being the probability of performing isomerization move. A sample of the correlation coefficients is shown in Figure 7. As we can see there, the correlation coefficients fall rapidly, thus requiring only a few Monte Carlo moves to get results with good accuracy. This is primarily due to the “all sublattice” optimization move through which a great variety of changes can be implemented at once on the current microstate.

The results of a sample statistics, that for the average number of vertical bonds entering a row from below, are compared next against the exact answers in Figure 8. As seen in the figure, there is indeed a very close agreement with the exact results (within 2 decimal places for the data shown in Figure 8 using just 50 million samples, less than 1 part per 240 of the total population). Note that these data were collected under unfavorable conditions for a Monte Carlo algorithm, i.e. in the absence of energetic effects where all states count equally thus necessitating a fairly broad and representative state sample to produce faithful estimates.

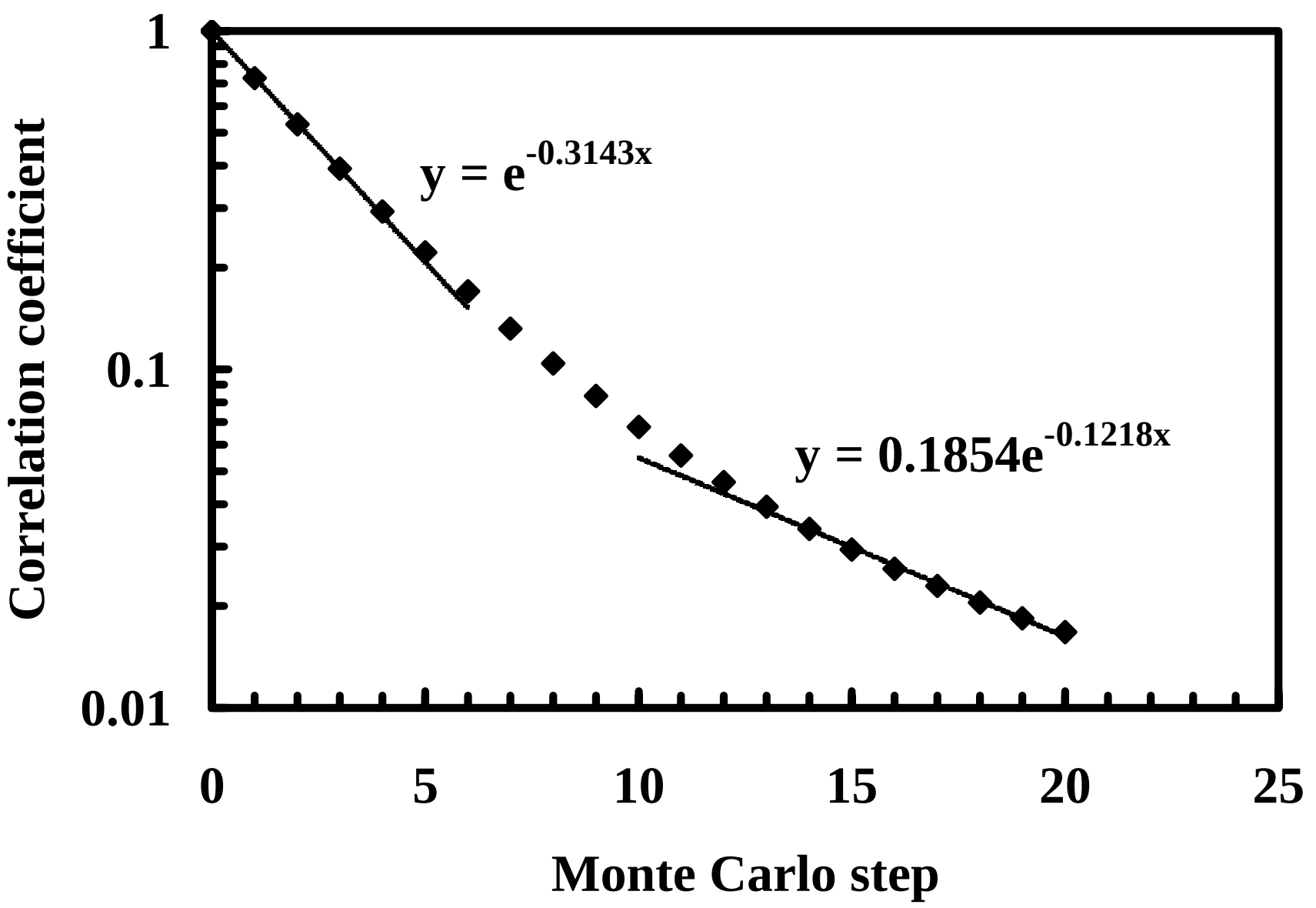

**Figure 7.** Correlation between Monte Carlo steps at zero energetics for an 8x8 node lattice. Note that the correlation between microstates at the end of specific number of Monte Carlo moves decays almost exponentially and has reached the a value of $e^{-1}$ (approx. 0.3) within 5 steps.

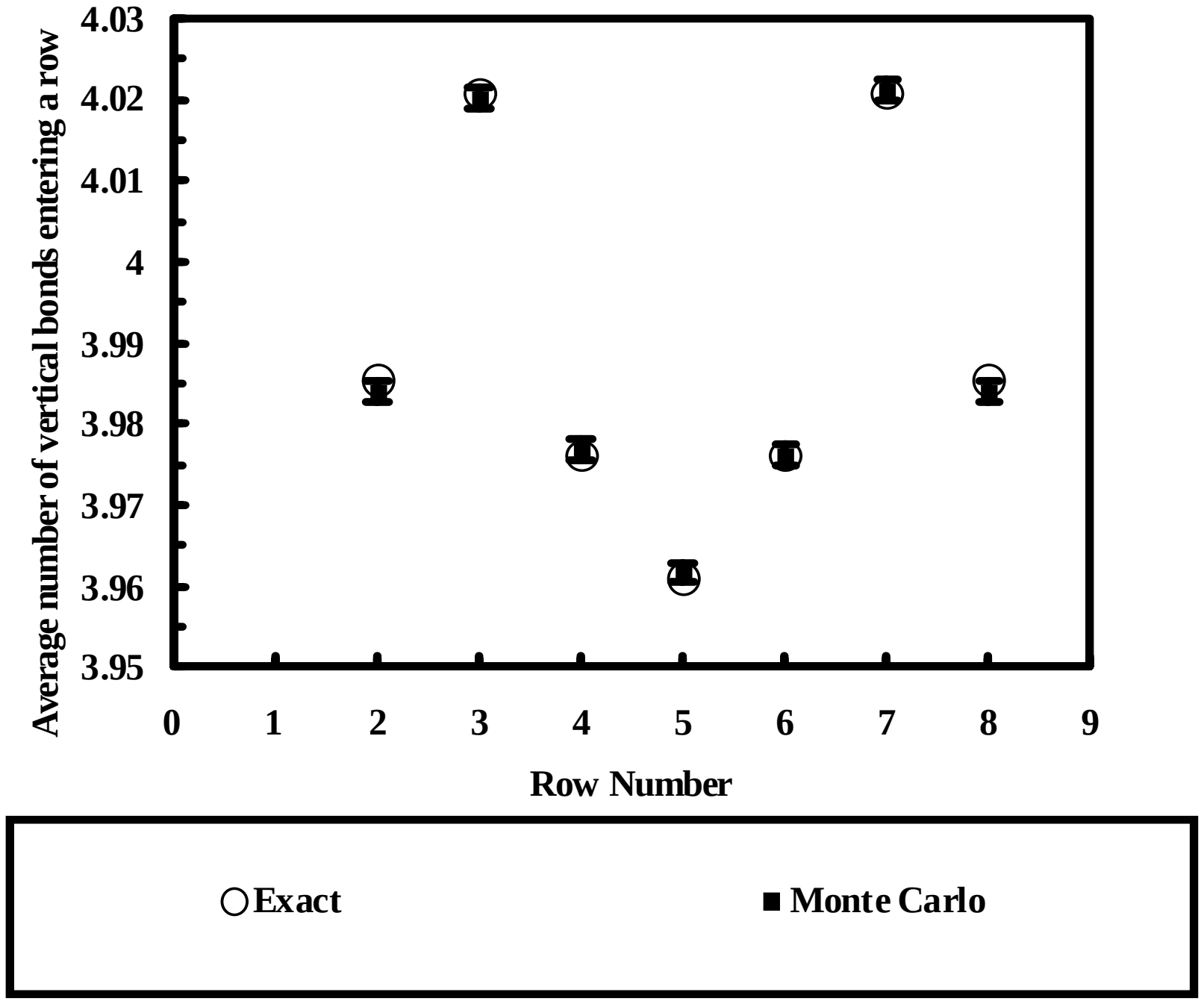

**Figure 8.** Comparison of the Monte Carlo (50 million samples) against the exact results from exhaustive enumeration for an 8x8 node lattice. The fraction of the vertical bonds entering a row from below is plotted against the row number. The Monte Carlo results show excellent agreement with the exact results with the latter always falling within the error window calculated a priori within the Monte Carlo simulations.

For the quantity of interest here (the conformation statistics for the average number of vertical bonds connecting a row of the lattice with the row below it) and for a 50 million Monte Carlo samples, the results were accurate to more than 3 significant digits. Of course the accuracy can increase by increasing the number of Monte Carlo moves performed: one can get a 4-digit accuracy with 1 billion Monte Carlo moves in about $1/10^{th}$ the time required for performing exhaustive enumeration of microstates approach. Moreover, the exact results fell always consistently, inside the error bars in all cases, with the error bars again calculated within the Monte Carlo scheme independently from the exact answers and indicate twice the estimate for the standard deviation.

Other chain statistics for the 8x8 node case this time, in the presence of energetic effects, are shown in Figure 9. These statistics are essential in order to understand how the thermodynamics and chain conformation change in the interlamellar amorphous region as a function of the physical parameters of the system: the height of the amorphous region and the energetic intra/inter chain interactions. Figure 9(a) shows the dependence of the fraction of chains forming tight folds in the first layer as a function of strain energetics. Figure 9 (b) and (c) show the dependence of the average number of bridges and loops, respectively, on the strain energetic parameter, $E_\eta$. In all the cases, the Monte Carlo results (obtained here with 200 to 800 million moves) are in very close agreement with the exact results. In fact the error bars are indistinguishable from the symbol in these figures. Furthermore, the influence of chain energetics can be clearly seen. As the energetic penalty (given by the dimensionless energy per fold parameter, $E_\eta$, non-dimensionalized with respect to Boltzmann factor, $k_BT$) on the tight folds structures increased, the fraction of chains that fold back to the same crystalline boundary in the very first layer (regular folds) quickly decreases (Figure 9 (a)). Simultaneously, the average number of bridges per microstate increases (Figure 9 (b)) and the average number of loops per microstate decreases (Figure 9 (c)). This behavior eventually results so that at higher energetic penalties per tight fold, approximately more than 5, the lattice is almost crystallized.

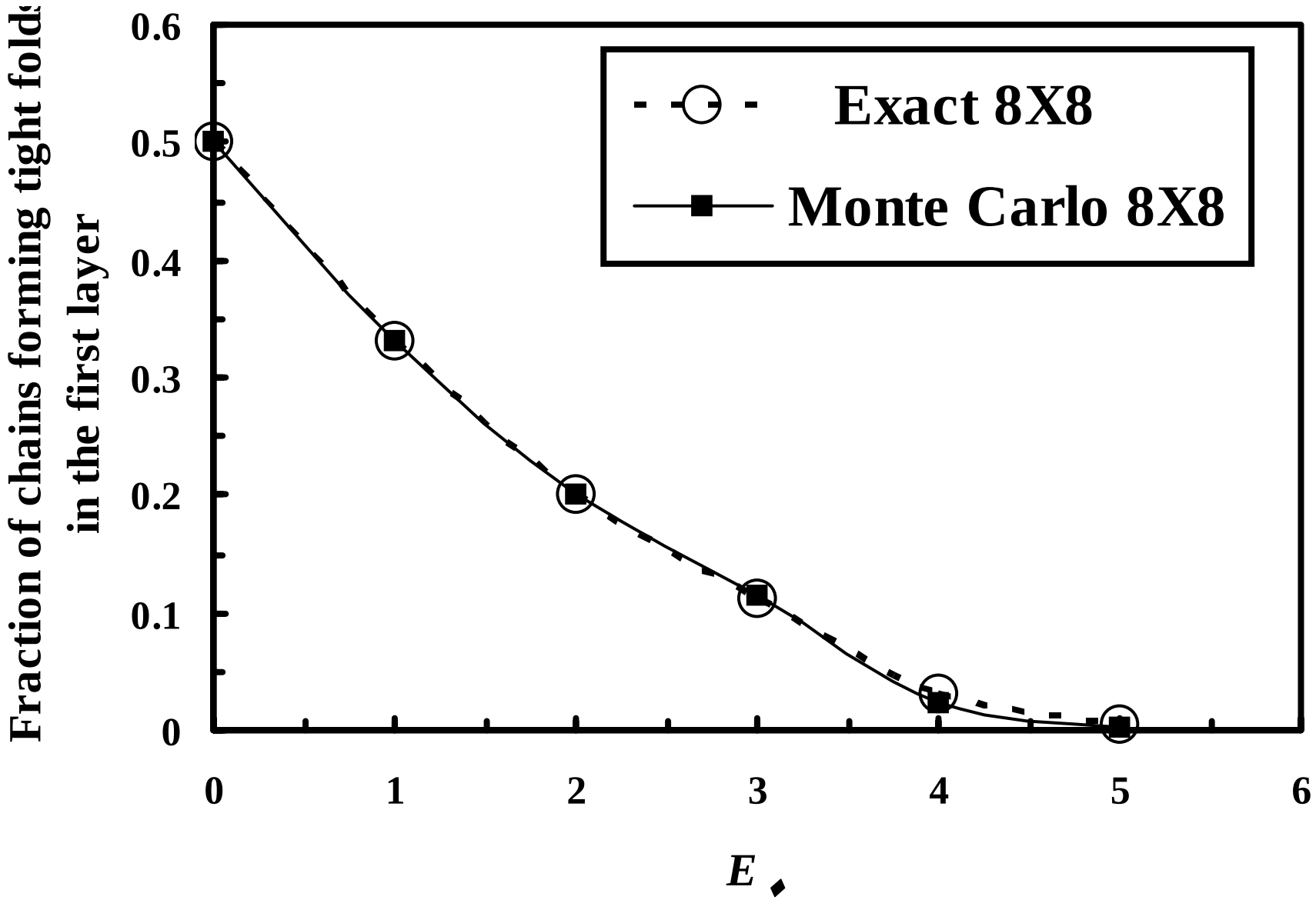

Figure 9 (a)

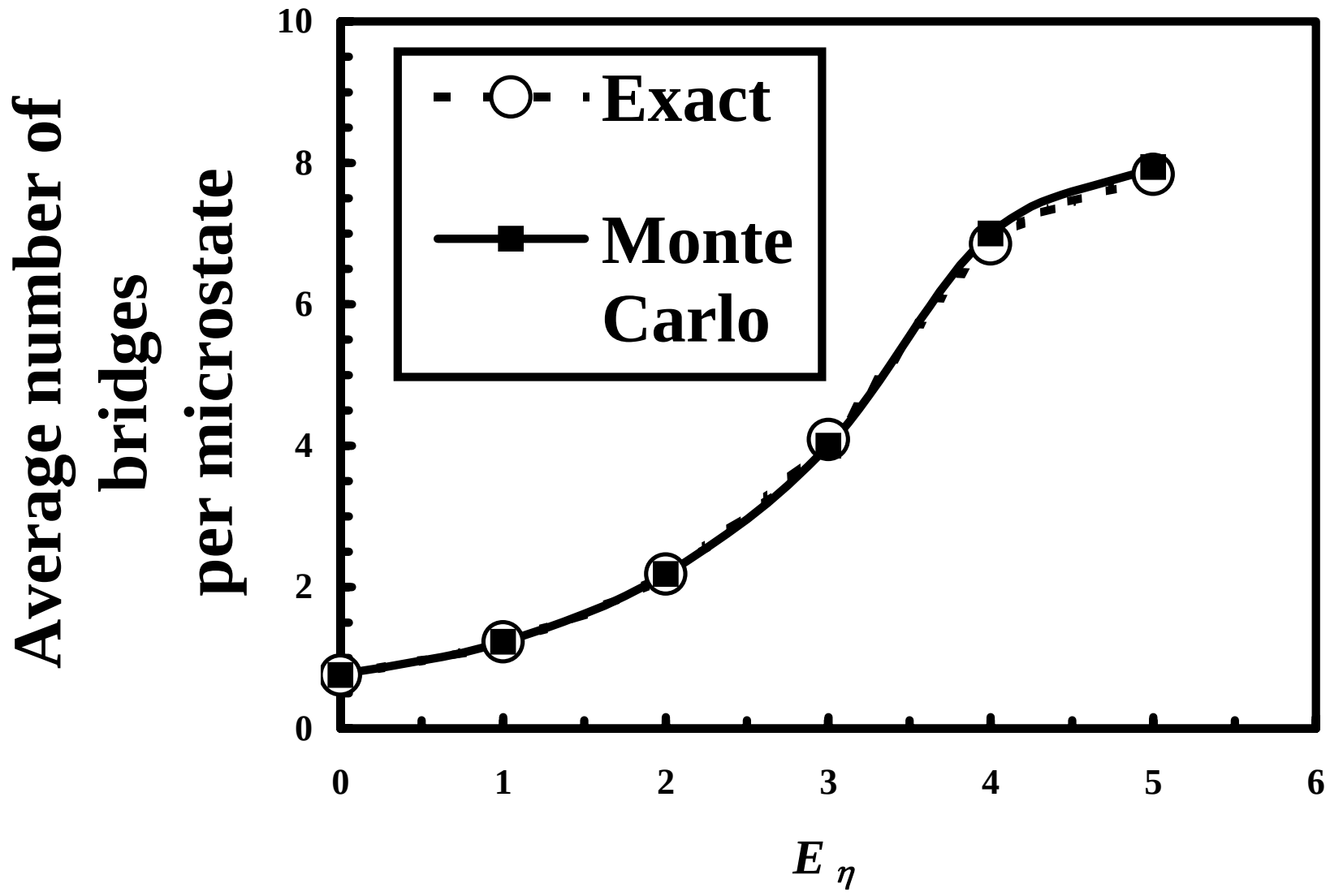

Figure 9 (b)

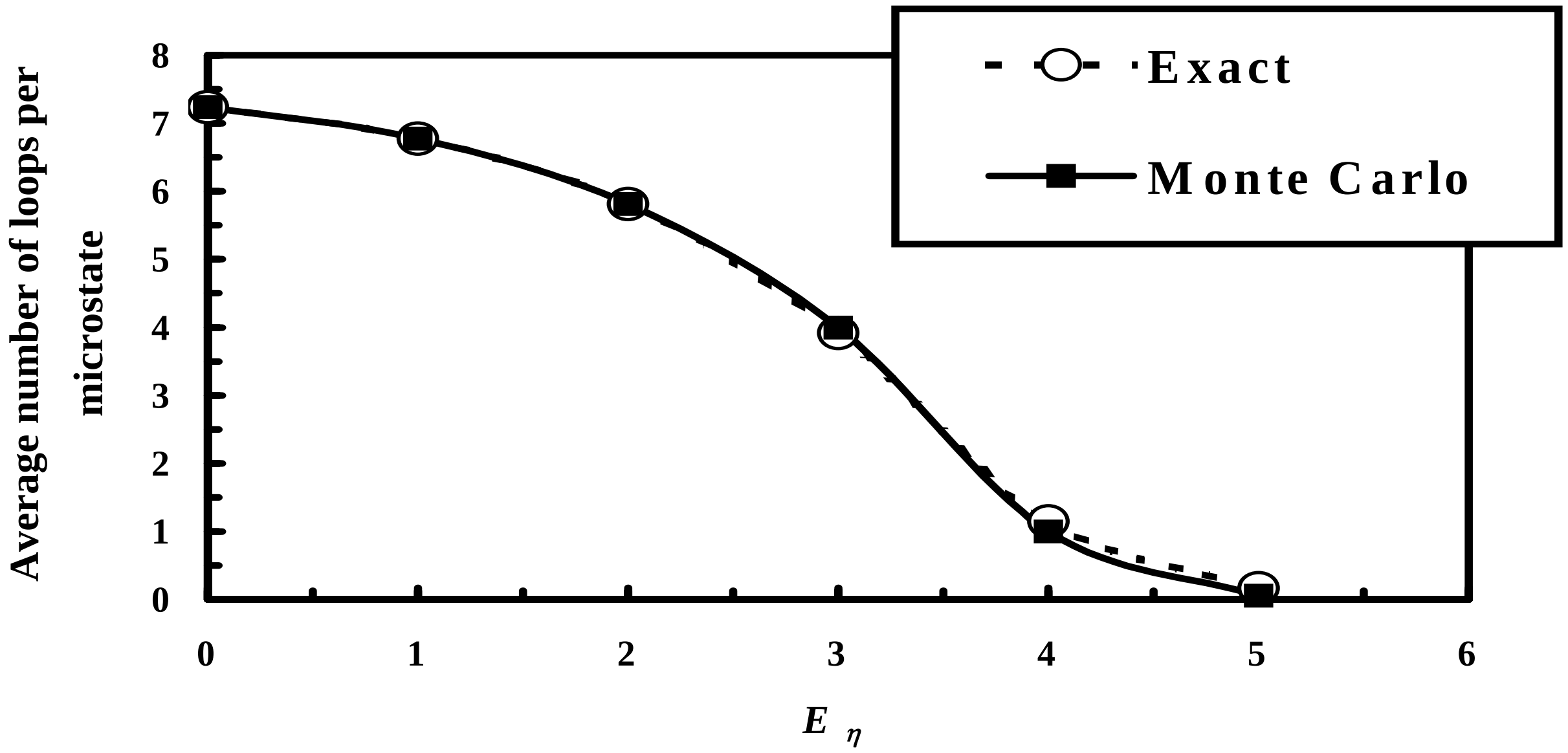

Figure 9 (c)

**Figure 9.** Comparison of the exact results from exhaustive enumeration to the approximate ones using Monte Carlo simulations (200 to 800 million samples per point) for an 8x8 node lattice. a) The fraction of chains forming tight folds in the first layer is plotted as a function of energy per fold parameter, $E_\eta$. b) The average number of bridges per microstate is plotted as a function of energy per fold parameter, $E_\eta$. c) The average number of loops per microstate is plotted as a function of energy per fold parameter, $E_\eta$.

A closer inspection of the Monte Carlo results and their comparison to the exact results leads here to the observation of two types of behavior. For small energetics, $E_\eta<3$ the exact results fell within independently calculated error bars (although those are too small to be seen within the scale of Figure 9). However, for high energies we saw the signs of a generic limitation of the Monte Carlo scheme i.e. the "local minimum trapping" of the solution next to a minimum, which even when is the global minimum, can result in systematic errors owed to the diminished representation of potentially more voluminous but energetically less favored states. This interpretation is consistent with the comparatively larger error seen at higher energetics (when the energy per fold is 4 and 5 - see Figure 9). The local energetic trapping problem can be circumvented by evaluating the various statistics corresponding to higher energetics using the Monte Carlo sampling at a lower energy and reweighting the statistics in a post-processing step[27]. The formula used to evaluate the reweighed statistics at high energies is shown in equation 12, where $E_b$ is the base energy and $E_c$ is the energy state at which the statistics are desired.

$$\overline{x_{E_c}} = \frac{\sum_{i=1}^{N_{sample}} x_{E_b,i} \exp\left(E_{b,i} - E_{c,i}\right)}{\sum_{i=1}^{N_{sample}} \exp\left(E_{b,i} - E_{c,i}\right)} \quad . \qquad \dots 12$$

Especially for our case, where the energetics are defined by tight fold structures only, we can effectively evaluate multiple statistics, pertaining to multiple values of $E_\eta$, using a single MC run at a reference (base) energy per fold value $E_b$. We accomplish this by first categorizing data for each possible given number of total tight folds in the MC generated microstructure, $i$, (which for an 8x16 node lattice studied here, can have a value ranging between 0 and 64) and then reweighting them as needed a posteriori to evaluate averages at higher eneregetics as shown in equation 13.

$$x_{E_{\eta c},i} = \frac{x_{E_{\eta b},i} \exp\left(\left(E_{\eta b} - E_{\eta c}\right) i\right)}{\sum_{i=0}^{32} x_{E_{\eta b},i} \exp\left(\left(E_{\eta b} - E_{\eta c}\right) i\right)} \quad . \qquad \dots 13$$

Here $x_{E,i}$ is the distribution of any quantity $x$ at an energy value of $E$, with respect to the number of folds, $i$.

Indeed when this is done, the above mentioned deficiencies disappear, as seen in Figure 10, where a comparison is made between the exact answers from the exhaustive enumeration scheme and those obtained from the reweighed Monte Carlo scheme. Further discussion on the effect of higher energetics and larger sizes, and extrapolation of results to the thermodynamic limit (i.e. for the lattice having infinite lateral extent) for the sake of brevity and organizational ease, will be deferred till the next part of this series[24], which discusses the application of the lattice subdivision methodology to larger lattice sizes.

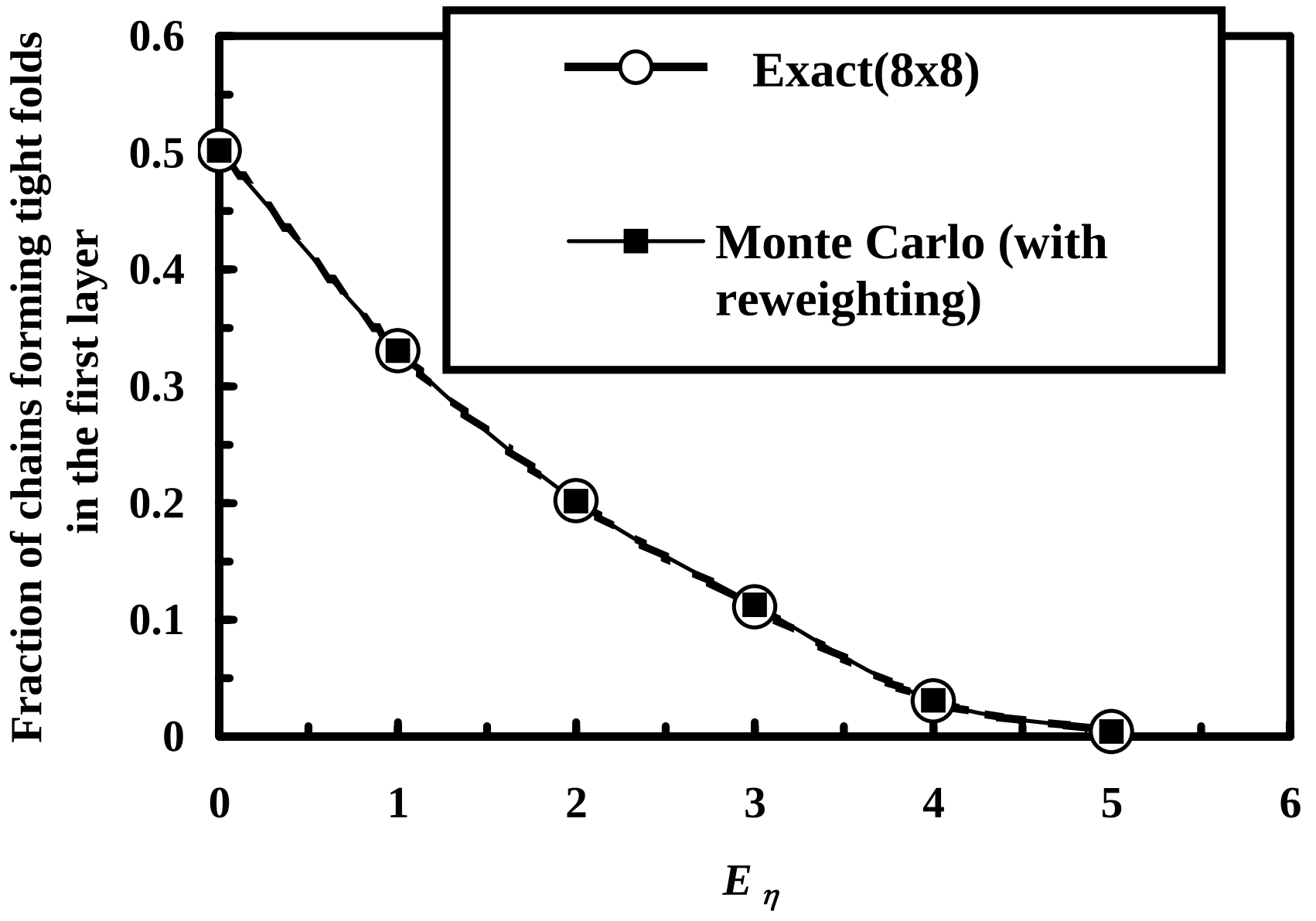

Figure10 (a)

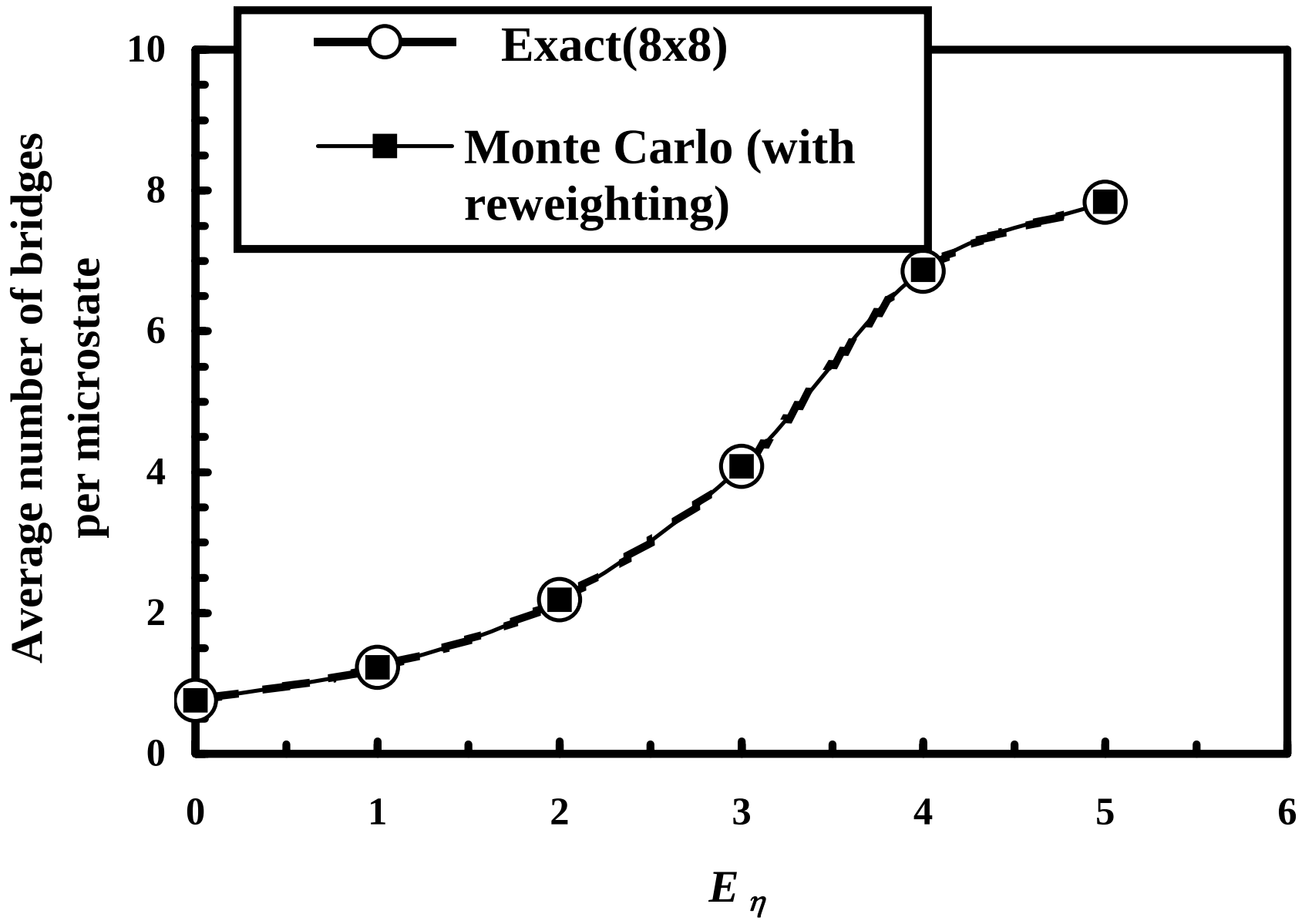

Figure 10 (b)

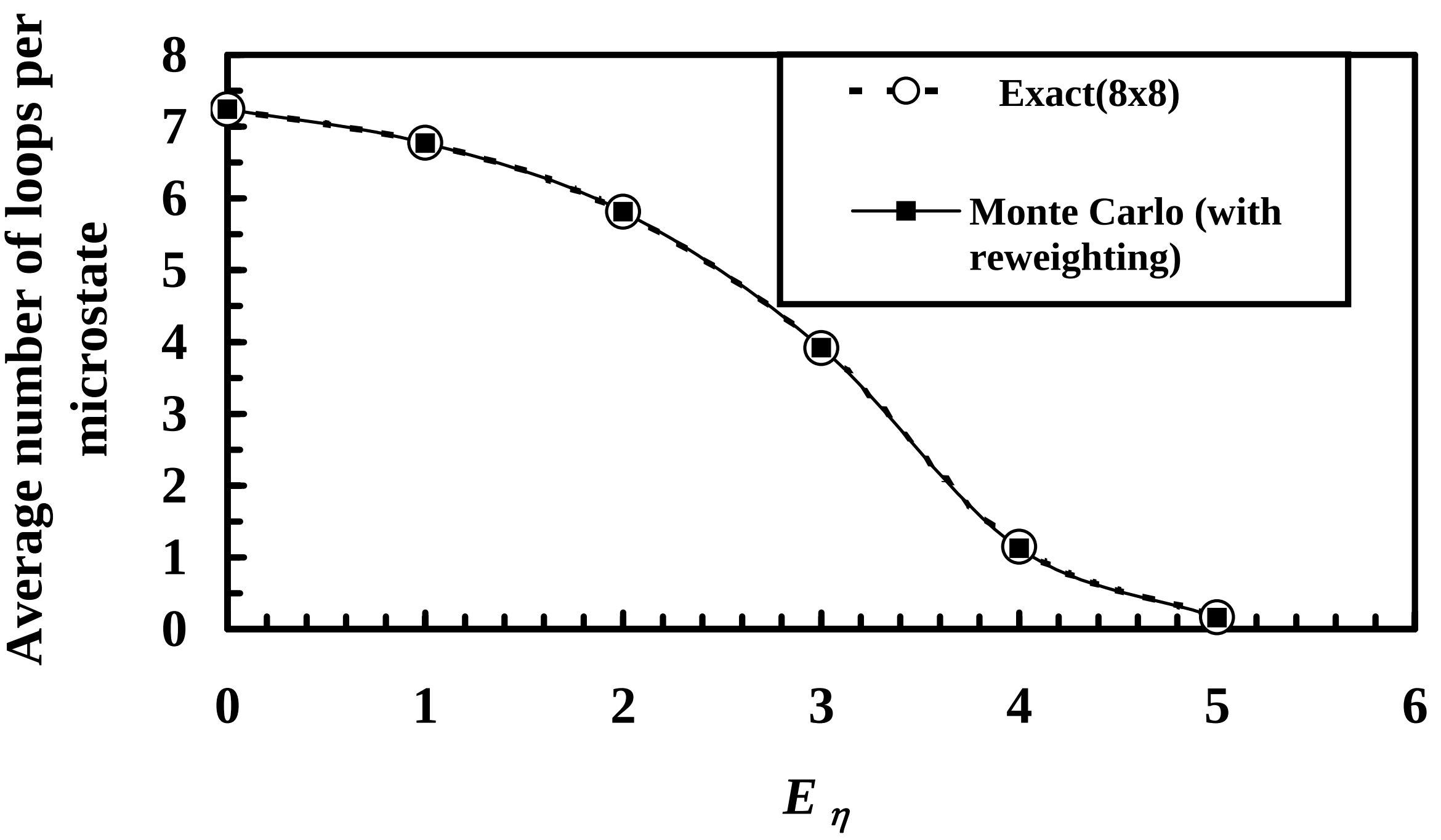

Figure 10 (c)

**Figure 10.** Comparison of the exact results from exhaustive enumeration to the results from Monte Carlo simulations (200 to 800 million samples per point) for 8x8 node lattice with results at $E_\eta$>2 obtained by reweighting the results from $E_\eta$=2. a) The fraction of polymer chains forming tight folds in the first layer is plotted as a function of the energy per fold parameter**,** $E_\eta$. b) The average number of bridges per microstate is plotted as a function of the energy per fold parameter**,** $E_\eta$. c) The average number of loops per microstate is plotted as a function of the energy per fold parameter**,** $E_\eta$.

## 5 Conclusions

In this work we developed and systematically validated new computational methods for a systematic investigation of the polymer chain conformations. Two different stochastic computational techniques have been used, all based in a lattice model of the polymer chains, illustrated here in two dimensions. As the common basis of all methods, a lattice subdivision technique has been developed which has been found to be very efficient. In the papers that follow, we will discuss application results in two-dimensional lattices[24] (especially in the thermodynamic limit) and an extension of the principles to three-dimensional lattices[25] respectively.

First, the exhaustive enumeration technique allowed us to obtain exact results for all the chain conformations in small sized lattices subject to specific connectivity conditions (boundary conditions). Since the applicability of the exhaustive enumeration technique is limited to small lattice sizes, the main application of this approach is for database information development and to validate stochastic approaches. The stochastic approaches allowed us to expand our investigations to substantially larger 2-D lattice sizes with or without energetic interactions. In particular the stochastic enumeration technique allows us to evaluate estimates of the absolute population size, which is of importance in determining absolute thermodynamic quantities. In addition, the use of a rich selection of moves within the Monte Carlo method leads to a very small correlation between subsequent Monte Carlo moves that allow for an efficient sampling of the population studied using fairly modest Monte Carlo sample sizes.

Moreover, simultaneous with the careful evaluation of the proposed techniques and with their significant computational efficiency, an important outcome of this work is the development of independently calculated measures of the error associated with the stochastic approaches. The predictions obtained from those approaches, with independently evaluated error bars, were found to always be consistent when compared against the exact results. This allows for the stochastic results to be used in a fully quantitative fashion. Extensive results obtained based on these techniques will be the subject of future publications.

**Acknowledgments**

The authors will like to acknowledge Prof. Sanat Kumar, formerly at Pennsylvania State University (now at Columbia University) and Dr. Chester Miller, formerly at the Experimental Station, Dupont, for their helpful suggestions and insightful comments. This work has been supported through a grant from National Science Foundation (award DMI 9978656 in materials processing and manufacturing). A separate NSF REU grant to the Department of Chemical Engineering, University of Delaware (for Timothy King) while he was an undergraduate student at University of Virginia is duly acknowledged.